# Self-Organized V – Mo Oxide Fibers by the Micro-Pulling Down Method

Detlef Klimm[1], Krzysztof Orlinski[2] and Dorota A. Pawlak[2]
[1]Leibniz Institute for Crystal Growth, Max-Born-Str. 2, 12489 Berlin, Germany.
[2]Institute of Electronic Materials Technology, 133 Wólczyńska Str., 01-919 Warsaw, Poland.

**ABSTRACT**

The $V_2O_5$–$MoO_3$ mixtures offer a whole range of materials where properties can be adjusted by simple modification of experimental parameters, which may be utilized for manufacturing metamaterials with on-demand properties. The $V_2O_5$–$MoO_3$ system contains an intermediate phase $V_9Mo_6O_{40}$, with a small fraction of $V^{4+}$ instead of $V^{5+}$. Consequently, this system should rather be considered as pseudobinary. The $V^{4+}$ content depends on the oxygen partial pressure in the atmosphere. Thus, by changing the oxygen partial pressure one can tailor the electric properties of the system, and by changing the supercooling, the morphologic structure of crystallized bodies as well. For better understanding of this system differential thermal analysis and thermodynamic modeling was performed. Fibers of eutectic composition between $V_9Mo_6O_{40}$ and $MoO_3$ were grown by the micro-pulling-down technique. X-ray diffraction confirmed the existence of the $V_9Mo_6O_{40}$ intermediate phase.

**INTRODUCTION**

In recent years special interest has been put to directional solidification of eutectic materials. Eutectic self-organization, which is a paradigm of pattern formation, has been proposed as potential method for manufacturing metamaterials while utilizing mechanisms of self organization [1,2,3]. Usually, eutectics are made or of two metals or of two insulating materials, and for such kind of materials there are the databases of phase diagrams, which are necessary to grow particular material. However considering the metamaterials application, it would be interesting to create self-organized patterns from components with highly different dielectric properties (e.g. from metal and insulator). This is not straightforward, as conventional metals (as conductors) do not form easily eutectics with oxides (as insulator) as a result of vanishing miscibility in the liquid state. However some metal oxides like e.g. $RuO_2$ show also high conductivity and might be useful to replace normal metals as a basic component of a metamaterial. Other metal oxides with high conductivity recently investigated are based on vanadium oxide and related mixed oxide systems [4]. The system $V_2O_5$–$MoO_3$ seems promising for the search for conductor/insulator eutectics, and some theoretical and experimental results are presented for the first time here.

Recently it was reported that nanostructured vanadium-molybdenum mixed oxides with different molar composition $x = [MoO_3]/([MoO_3]+[V_2O_5])$ possess electronic conductivity σ, especially for medium concentrations [4]. It was found that σ is related to $V^{4+}$ ions, that occur within all samples together with $V^{5+}$, except for the pure components molybdenum (VI) oxide $MoO_3$ and vanadium (V) oxide $V_2O_5$. Besides for electric conductivity, the charge transfer between the different vanadium ions is considered to be responsible for the catalytic activity of V-Mo oxides [5]. Already Magnéli [6] pointed out that the crystal structures of $V_2O_5$ and $MoO_3$ are somewhat similar, as both can be considered to be built up from $MeO_6$ octahedra. About 17% of the $V^{5+}$ ions in $V_2O_5$ can be substituted by $Mo^{6+}$, and the charge balance can the easiest be obtained by partial reduction of vanadium to $V^{4+}$. This limit corresponds to a molar solubility $x = 0.25$. For larger $x$, an intermediate phase with the approximate composition $V_2MoO_8$ was found.

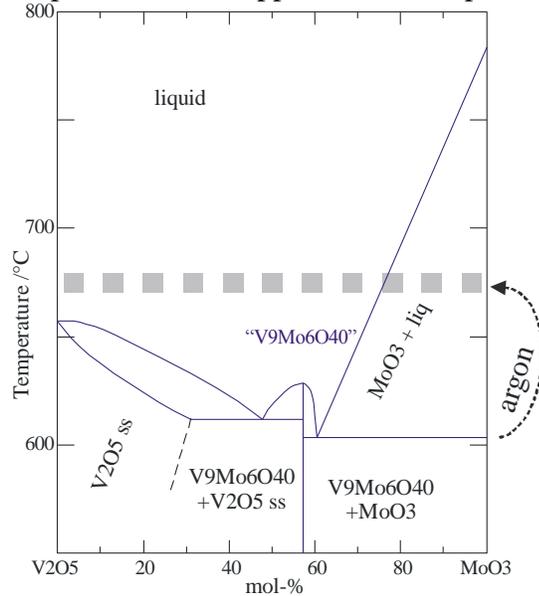

Fig. 1: Blue solid lines: The phase diagram $V_2O_5$–$MoO_3$ as proposed by Bielanski et al. [7] was basically confirmed by the current DTA measurements in air. If the measurements are performed in argon, one eutectic is observed at $(674\pm3)°C$ instead (broad dashed line).

Later Bielanski et al. [7] reported a version of the $V_2O_5$–$MoO_3$ phase diagram, that is shown by the solid lines in Fig. 1, with $V_9Mo_6O_{40}$ ($x = 0.5714$, fusion point $T_f = 628°C \approx 900$ K) as intermediate compound. It must be noted, however, that this is not a real binary phase diagram, as the intermediate phase is not a linear combination of the two end members:

$$9\ V_2O_5 + 12\ MoO_3 = V_{18}Mo_{12}O_{81} \neq 2\ V_9Mo_6O_{40} \tag{1}$$

In this work, simultaneous differential thermal analysis (DTA) and thermogravimetric (TG) measurements were performed under different atmosphere to reveal relevant phase relations in the system $V^{5+}{}_2O_5$–$V^{4+}O_2$–$Mo^{6+}O_3$. Besides, first eutectic fibers were grown on the $MoO_3$ rich side to check the properties of such system that is built up from a phase with low σ ($MoO_3$) and a phase where σ is high ($V_9Mo_6O_{40}$).

**EXPERIMENT**

Most thermoanalytic measurements where performed with a NETZSCH STA 449C "Jupiter" and with double DTA/TG sample carrier, where two "type S" thermocouples for sample and reference are serial connected leading to higher sensitivity compared with standard carriers. A vacuum-tight rhodium DSC furnace allowed the accurate control of atmosphere that could be switched during the DTA/TG runs between air and argon with 99.999% purity. For some measurements in flowing air a NETZSCH STA409C was used. Samples of 20–30 mg in lidded Pt crucibles where measured with heating/cooling rates of ±10 K/min up to a maximum temperature of $T_{max}=800°C$, that was used only for pure $MoO_3$. For all other samples, that were obtained by mixing appropriate amounts of 99.9% purity $V_2O_5$ (Riedel de Haën) and 99.998% purity $MoO_3$ (Alfa Aesar), the liquidus was lower and $T_{max}\approx700°C$ allowed to keep evaporation loss of the volatile oxides negligible. The thermoanalytic setup was calibrated for $T$ and sensitivity at the melting points of zinc and gold, and at the phase transformation point of barium carbonate. Fig. 2 and Fig. 3 show DTA and TG curves for some samples. Second or later heating/cooling curves were used for analysis to ensure good homogenization of the samples that was obtained during a first heating/cooling cycle.

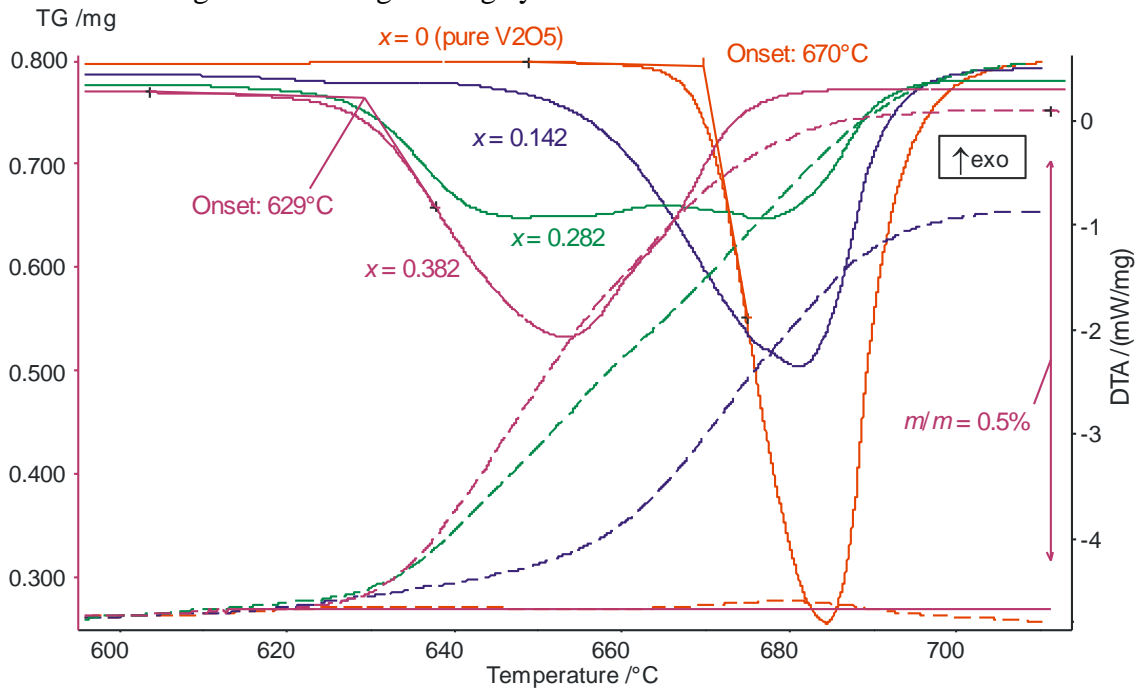

**Fig. 2: DSC (solid) and TG (dashed) of second heating runs for four samples from pure $V_2O_5$ to $V_2O_5$–$MoO_3$ with $x = 0.382$ in air.**

As expected, pure $V_2O_5$ ($x = 0$, red curve in Fig. 2) shows one sharp melting peak with extrapolated onset at 670°C, which is in perfect agreement with literature data [9]. For measurements in air (Fig. 2) as well as argon the $V_2O_5$ mass remains constant during the melting process. If $MoO_3$ is added, the melting onset drops down to almost constant values <630°C for $x>0.3$, besides the melting peak becomes substantially broader. Summarizing it was found that the DTA curves from measurements in air could very well be described by the phase diagram that is shown above (Fig. 1). From the melting peak area at the composition $V_9Mo_6O_{40}$ the heat of fusion $\Delta H_f \approx (384\pm40)$ J/g = 642,854 J/mol was measured for this intermediate phase.

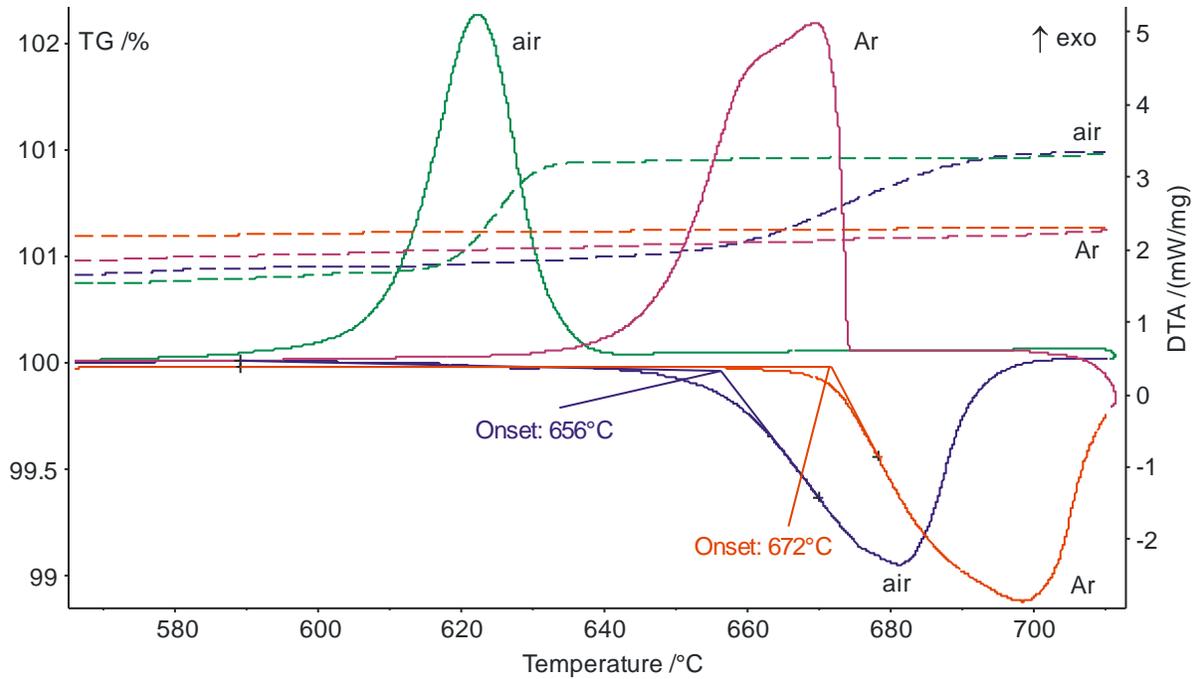

**Fig. 3: DSC (solid) and TG (dashed) of second heating/cooling runs for a $V_2O_5$–$MoO_3$ sample with $x = 0.142$ in air and in argon (99.999% purity).**

Fig. 3 shows that the DTA and TG curves for one and the same sample are completely different, if the measurements are performed in flowing air or in flowing argon, respectively. The residual impurities in the commercial gas result in an oxygen partial pressure $p_{O2}$ of several $10^{-6}$ bar [8]. It is remarkable that only in Ar the sample mass $m$ remains constant during melting and solidification, whereas $m$ grows as a result of the uptake of oxygen during melting in air, and $m$ drops reversibly during solidification in air. Besides, the melting onset of the mixture that is shown in Fig. 3 shifts upwards by 16 K: melting and solidification peaks are starting now at $T \approx 675°C$.

The raw material with 60mol% $MoO_3$ and 40mol% $V_2O_5$ (this is the eutectic composition on the $MoO_3$ rich side in Fig. 1) was prepared as starting material for µ-PD growth experiments, and eutectic rods were obtained in air with a pulling rate of 1.5 mm/min. During growth the formation of gas was observed at the crystal/melt interface. Thus the gas was trapped inside the growing rod, making the material porous. By lowering the pulling rate to less than 1 mm/min the growth of fibers with low fraction of cavities was possible. It should be noted that the gas release during crystallization is in agreement with the TG mass drop that is shown in Fig. 3 (green dashed line).

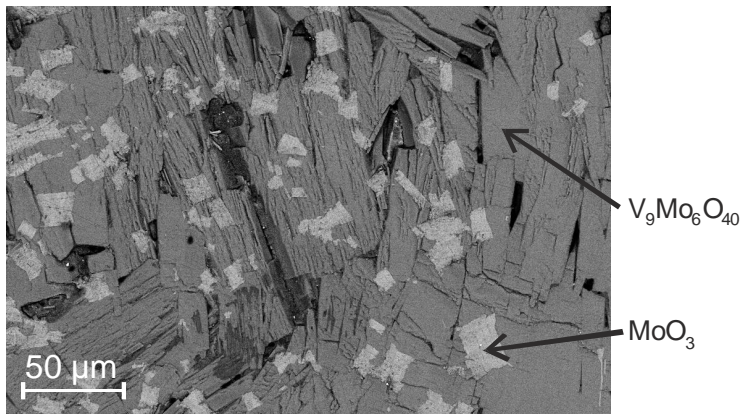

**Fig. 4: SEM micrograph (backscattered electrons) of a eutectic fiber showing MoO$_3$ inclusions in a V$_9$Mo$_6$O$_{40}$ matrix.**

The fibers were characterized by Scanning Electron Microscopy (SEM) with EDX chemical analysis and by X-ray diffraction (Siemens D500 using Cu-Kα radiation). The first growth experiments reveal a microstructure of MoO$_3$ rectangular precipitates embedded in the V$_9$Mo$_6$O$_{40}$ matrix (Fig. 4). Unfortunately, a phase with low σ (MoO$_3$) is embedded here in a mother phase with larger, but adjustable σ (V$_9$Mo$_6$O$_{40}$ with variable V$^{4+}$/V$^{5+}$ ratio). So far, a useful application for this self-organized structure was not found. Although the EDX spectra showed three different levels of V:Mo:O concentrations, only the two expected phases (intermediate phase and molybdenum oxide) where found in the X-ray patterns (Fig. 5).

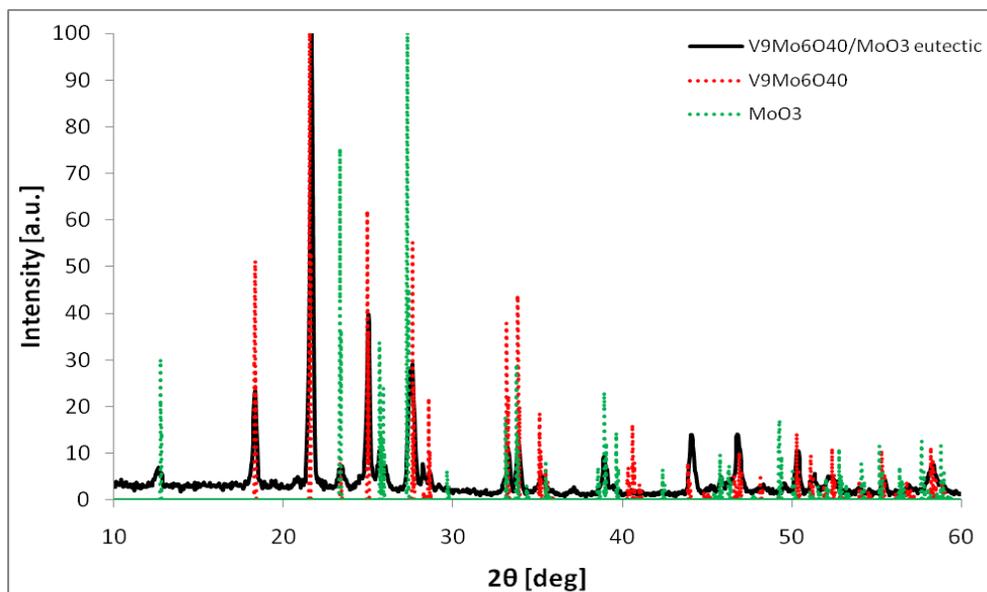

**Fig. 5: The X-ray diffraction pattern of a rod grown from 60% MoO$_3$ and 40%V$_2$O$_5$ eutectic composition $x$ = 0.6 (Fig. 1) showing only the expected V$_9$Mo$_6$O$_{40}$ and MoO$_3$ peaks.**

## DISCUSSION

For a quantitative description of the system, phase equilibria were calculated with the Gibbs energy $G$ minimization program FactSage [9]. Required $G(T)$ data for the relevant V-Mo-O species were found in the FactSage databases, except for the intermediate V$_9$Mo$_6$O$_{40}$. The following procedure gives an estimation for $G(T)$ for this phase: The atomic environments in the mixed oxide are comparable with the simple component oxides and formally one can write

$$6 \text{ MoO}_3 + \tfrac{1}{2} \text{ V}_2\text{O}_4 + 4 \text{ V}_2\text{O}_5 \rightarrow \text{V}_9\text{Mo}_6\text{O}_{40} + \Delta H \quad (2)$$

where the formation energy $\Delta H$ should be negative as the intermediate phase was really formed. For the starting materials in equation (2) one finds the enthalpies for solid and liquid state at 900 K (the melting point of $\text{V}_9\text{Mo}_6\text{O}_{40}$) that are given in Table 1. From these data one calculates together with the measured $\Delta H_f = 642{,}854$ J/mol

$$\Delta H = 6 \times 45{,}315 \text{ J/mol} + \tfrac{1}{2} \times 86{,}220 \text{ J/mol} + 4 \times 64{,}120 \text{ J/mol} - 642{,}854 \text{ J/mol}$$
$$\Delta H = -71{,}374 \text{ J/mol} \quad (3)$$

for the heat of $\text{V}_9\text{Mo}_6\text{O}_{40}$ formation to be entered in equation (2).

**Table 1: Enthalpies (in J/mol) of starting materials in equation (2) at 900 K [9]**

|  | $\text{MoO}_3$ | $\text{V}_2\text{O}_4$ | $\text{V}_2\text{O}_5$ |
|---|---|---|---|
| $H_{sol}$ (solid) | -690,146 | -1330600 | -1451410 |
| $H_{liq}$ (liquid) | -644,831 | -1244380 | -1387290 |
| $H_{liq} - H_{sol}$ | 45,315 | 86,220 | 64,120 |
| factor from (2) | 6 | ½ | 4 |

The "mixer" module of FactSage supplied a first estimation of the thermodynamic data of $\text{V}_9\text{Mo}_6\text{O}_{40}$, and this estimation was completed by the $\Delta H$ value from (3). Table 2 shows the resulting Gibbs energy data for the intermediate phase between room temperature and melting point.

**Table 2: Estimated Gibbs energy $G(T)$ of $\text{V}_9\text{Mo}_6\text{O}_{40}$**

| $T$ (K) | 300 | 400 | 500 | 600 | 700 | 800 | 900 |
|---|---|---|---|---|---|---|---|
| $G$ (kJ/mol) | -11812.2 | -11933.2 | -12083.4 | -12258.4 | -12454.9 | -12670.5 | -12903.2 |

Both $\text{MoO}_3$ and $\text{V}_2\text{O}_5$ have low melting points below 800°C or 700°C, respectively. Contrary for the third component oxide of the intermediate phase, $\text{V}_2\text{O}_4$, a very high $T_f \gg 1300°C$ is reported by all authors (FactSage [9] reports 1545°C).

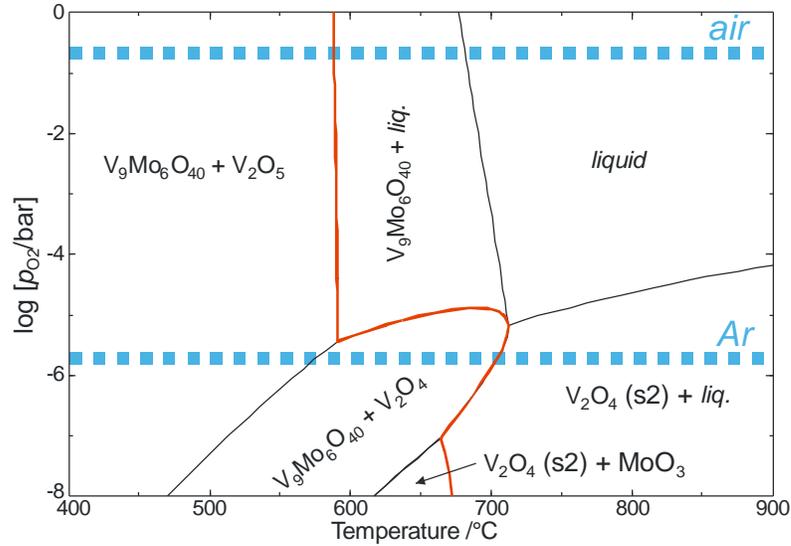

**Fig. 6: Phase fields of vanadium-molybdenum-oxides with V:Mo in slight excess of the ideal 9:6 ratio for $V_9Mo_6O_{40}$. The liquid phase of molten V-Mo-oxides is calculated as ideal mixture. The dashed lines represent the $p_{O2}$ of air or of the residual gas impurities in Ar with 99.999% purity, respectively.**

Fig. 6 shows calculated phase fields of V-Mo-oxides that are based on literature data [9] for all phases – except for $V_9Mo_6O_{40}$, where the data from this work (Table 2) have been used. The thick red line, starting from the top near 600°C, marks the melting process, as only right from this line phase fields containing *liquid* are found. The calculated phase diagram underestimates the melting onset of $V_9Mo_6O_{40}$ in air by ca. 40 K. This error could be due to the simplifying assumption of an ideal melt (*liquid*), which had to be taken as data on the excess enthalpy of this melt are missing. Nevertheless, the significantly different behavior of V-Mo-oxides in different atmosphere (Fig. 3) can be explained at least semi-quantitatively: Only in air, $V^{5+}$ is stabilized to a high degree and $V_9Mo_6O_{40}$ is in equilibrium with the low melting $V_2O_5$. Contrary, in Ar under low $p_{O2}$, $V^{4+}$ is in equilibrium and leads to a significantly higher melting point.

## CONCLUSIONS

It could be shown that the 60% $MoO_3$ eutectic of the system $V_2O_5$–$MoO_3$ is suitable for the growth of fibers using the micro-pulling-down technique with pulling rates ranging from 1 to 5 mm/min. During the solidification process oxygen is released, resulting from the partial reduction of $V^{5+}$ to $V^{4+}$. With low pulling rate the gas is leaving the melt before entire crystallization, thus enabling the creation of bulk crystalline fiber. The microstructure is built from two phases: $V_9Mo_6O_{40}$ (containing multivalent vanadium) and $MoO_3$. EDX and XRD measurements show that molybdenum trioxide phase should be considered as a solid solution, as the peaks are shifted to lower angles, which might be caused by smaller vanadium atoms, replacing molybdenum and thus lowering cell parameters.

ACKNOWLEDGMENTS

The authors express their gratitude to the European Union for financial support in the 7[th] Framework Programme ("ENSEMBLE", NMP4-SL-2008-213669), KO and DAP thank the Project operated within the Foundation for Polish Science, Team Programme, co-financed by the EU European Regional Development Fund for support.